\documentclass[twocolumn,showpacs,amsmath,amssymb,pre]{revtex4}

\usepackage{amsthm}
\usepackage{MnSymbol}
\usepackage{graphicx}
\usepackage{dcolumn}
\usepackage{bm}
\usepackage{chemarrow}
\usepackage{graphicx}
\usepackage{epsfig}
\usepackage{mathrsfs}
\usepackage{nicefrac}

\newcommand{\be}{\begin{equation}}
\newcommand{\ee}{\end{equation}}
\newcommand{\bea}{\begin{eqnarray}}
\newcommand{\ba}{\begin{array}}
\newcommand{\eea}{\end{eqnarray}}
\newcommand{\ea}{\end{array}}

\begin{document}

\newtheorem{theorem}{Theorem}

\title{Macroscopic constraints for the minimum entropy production principle}

\author{Matteo Polettini}
 \email{polettini@bo.infn.it}
\affiliation{Dipartimento di Fisica, Universit\`a di Bologna, via Irnerio 46, 40126 Bologna, Italy\\
 INFN, Sezione di Bologna, via Irnerio 46, 40126 Bologna, Italy}

\date{\today}

\begin{abstract}
In an essential and quite general setup, based on networks, we identify Schnakenberg's observables as the constraints that prevent a system from relaxing to equilibrium, showing that, in the linear regime, steady states satisfy a minimum entropy production principle. 
The result is applied to master equation systems, opening a new path to a well-known version of the principle regarding invariant states. Moreover, with the aid of a simple example, the principle is shown to conform to Prigogine's original formulation. Finally, we discuss analogies and differences with a recently proposed maximum entropy production principle.
\end{abstract}


\maketitle

\subsection{Introduction}

The \textit{minimum entropy production principle} (MINEP) asserts, using Klein and Meijer's words \cite{klein}, that
\begin{quote}
\textquotedblleft the steady state is that state in which the rate of entropy production has the minimum value consistent with the external constraints which prevent the system from reaching equilibrium\textquotedblright .
\end{quote}
So worded, it is reminiscent of the inferential method that provides ensembles in  equilibrium statistical mechanics, by maximization of the Gibbs-Shannon entropy --- a measure of ignorance of the microstate of the system --- under suitable constraints. In an information-theoretic framework, constraints are pieces of knowledge the observer gains from the measurement of certain observables of the system, or macroscopic parameters that can be experimentally controlled. In the extremization procedure constraints are introduced through Lagrange multipliers \cite{maxent1}.

Not exactly so for MINEP. Its first proof as a closed theorem is attributed to Prigogine \cite{prigo}. In Prigogine's statement, owing to the applied thermo-chemical setting, knowledge of the nonequilibrium external constraints, such as temperature or chemical potential gradients, is granted from the start. Further generalizations of the principle always entail that constraints can be read off the physical setup of the problem. So, in his extension of the principle to density matrices \cite{callen}, Callen  recalls that
\begin{quote}
\textquotedblleft Prigogine showed that in the steady state which is reached when certain affinities are constrained to have definite values, all unconstrained affinities assume the values which minimize the entropy production function\textquotedblright .
\end{quote}
However, the environmental  influence on a system might be \textit{a priori} unknown, or difficult to decipher. In this paper we do not assume previous knowledge of the nonequilibrium constraints, or affinities. As uninformed observers, we look at the system, measure its fluxes and local constitutive relations, and ask which are the constraints that impede relaxation to equilibrium.

\subsubsection{Scope and plan of the paper} 

More specifically, this work addresses two technical questions: Which Lagrange multipliers should be introduced in the MINEP variational procedure? Which constraints are implicitly encoded in the transition rates of master equation systems? For systems in the linear regime, the answer is found in Schnakenberg's theory of macroscopic observables \cite{schnak}, further supporting the point of view that his construction identifies the fundamental, experimentally accessible observables of NESM (Non-Equilibrium Statistical Mechanics).
When the linear constitutive relations or probabilistic transition rates are known or measurable, Schnakenberg's affinities can be explicitly calculated. For chemical reaction networks, it is known that they coincide with chemical potential differences \cite{andrieux}. For this reason his analysis has mainly found application in biochemistry \cite{hill,schnak2}, where it plays an important role for the comprehension of free energy transduction. Recently it is finding growing applications to transport phenomena and molecular motors \cite{andrieux,liepelt,faggionato,qians,seifert2}. These works show how the seemingly rather formal theory makes direct contact with experimentally accessible problems in thermodynamics, such as bounding maximum power efficiencies of nanomachines \cite{seifert2}.

We first identifty the underlying degrees of freedom that are subject to constraints,  spoiling the problem of its material content and considering bare fluxes of ``information'' to achieve the generality of equilibrium statistical mechanics \cite{maxent1}. For systems consisting of a finite number of microstates, we identify Schnakenberg's  affinities \cite{schnak} as the correct macroscopic constraints. Schnakenberg  introduced them for Markovian systems whose evolution is dictated by a master equation, but the construction can be generalized to any network of currents. Affinities are  defined as circuitations of nonconservative forces. Along with their conjugate ``mesh'' currents, they furnish a complete description of the steady state.

As for most, if not all, constructive variational principles in NESM, the range of validity of the principle is the linear regime. Notice that we will \textit{assume} the linear regime constitutive relations and not derive them from the principle of least dissipation, as is done in classical  textbooks on nonequilibrium thermodynamics \cite[Ch. 4 and Ch. 5]{gyarmati}. In particular, we will not distinguish between the dissipation function and the entropy production.

Schnakenberg's network theory can be efficiently introduced in an algebraic graph-theoretical fashion. We will concisely  provide all the necessary tools in Sec.\ref{schnak}, in a self-contained manner. For more details, good references are Biggs's book \cite{biggs} and Nakanishi's \cite{nakanishi}. The principle is formulated in Sec.\ref{MINEP}, and then applied to Markovian master equation systems. This yields a proof to the fact that the steady state is a local minimum of the entropy production. We thus obtain by a very different method a result previously derived by Jiu-Li, Van den Broeck and Nicolis \cite{luo}, and more recently rediscovered in the framework of Large Deviation Theory by Maes and Neto\v cn\'y \cite{maes1}. These earlier results are discussed in Sec.\ref{relationship}, where the relationship with Prigogine's original statement of the principle is also discussed. In fact, it turns out that the principle is perfectly adherent to Prigogine's formulation. The last subsection of Sec.\ref{relationship} is devoted to a comparison of our result with a formulation of the maximum entropy production principle due to  P. \v Zupanovi\'c, D. Jureti\'c and S. Botri\'c  \cite{looplaw}, which also deals with conservation laws, and might appear to be in contradiction --- at least nominally --- with ours, showing that the two principles are compatible.

We prosecute this introduction with a simple physical example that should convey that circuitations are good nonequilibrium constraints.

\subsubsection{\label{example} Circuitations as constraints} 

Consider the classical problem of heat diffusion in an approximately one-dimensional inhomogenous conductive rod, whose ends are put in contact with thermal reservoirs at slightly different boundary temperatures, $ T_b \gtrsim T_a$, while the body of the rod is isolated (see \cite[\S 3.1]{martyu2} and references therein). A temperature profile $T(x)$ establishes. 
By Fourier's Law the induced heat current thorugh the rod is
\be j(x) =  - k(x) \partial T(x),  \ee
where $k(x)$ is the thermal conductivity at $x \in [a,b]$ and $\partial = \nicefrac{\partial}{\partial x}$. The following identity
\be T_b - T_a + \int_a^b 	\frac{j(x)}{k(x)} dx \; \equiv \; 0   \label{eq:cond} \ee
is interpreted as a constraint on the currents, where we make use of the equivalence symbol `$ \equiv $' to impose constraints. Independently of the particular evolution equation that the heat current satisfies, the configuration is said to be \textit{steady} if at each point of the rod the influx of current balances the outflux, $\partial j^\ast(x) = 0$, which implies $j^\ast(x) = const$.

The same result can be obtained by a different route. Let us define the \textit{local force} as the right incremental ratio
\be a(x) = \lim_{\delta x \to 0^+} \frac{T^{-1}(x+\delta x) - T^{-1}(x)}{\delta x},  \quad x \in [a,b) . \ee
We assume that the system satisfies linear regime constitutive equations, that is, that forces and currents are small and  linearly related,
\be
a(x) = \ell(x) j(x). \label{eq:linregcon} 
\ee
For this assumption to hold it is necessary that the temperature profile is approximately constant and the temperature drop between the extremities of the rod is sufficiently small with respect to its length, in such a way that to first order one can approximate the conjugate force as $a(x) = - T_{av}^{-2} \partial T (x)$, $x \neq b$, where $T_{av}$ is the average value of the temperature \cite{debate2,martyu2}. The local linear regime coefficient then reads $\ell(x) = T_{av}^{-2} k(x)^{-1}$.
The entropy production
\be \sigma = \int_a^b j(x)a(x) dx, \ee
is then a quadratic functional of the currents. We require $\sigma$ to be stationary, that is $\delta \sigma =0$, with respect to all possible current profiles that are consistent with constraint (\ref{eq:cond}). Introducing one Lagrange multiplier $\lambda$, we calculate the variation
\be \frac{\delta}{\delta j(y)}\left[ \int_a^b \ell(x) j(x)^2 dx - 2\lambda   \int_a^b 	\frac{j(x)}{k(x)}  dx\right] = 0, \ee
leading to a uniform stationary current $j^\ast = \lambda   T_{av}^{2}$.
The value of the multiplier is fixed by substitution into Eq.(\ref{eq:cond}):
\be
j^\ast = T_{av}^2 \lambda = (T_b - T_a) \left[ \int_a^b k(x)^{-1} dx \right]^{-1} .
\ee
The above solution corresponds to a minimum of $\sigma$, and it coincides with the steady configuration of currents. We conclude that the steady state is the minimum of the entropy production among nearby current profiles that are compatible with the external constraint. Notice that, while in the NESM literature ``stationary'' and ``steady'' are synonyms, we prefer to use the former when referring to the extremal solution of a variational problem, and the latter for a configuration of currents that satisfies the continuity equation. 

The problem of heat-conduction and the minimum entropy production principle in a rod has been widely debated \cite{debate,debate2}, with arguments revolving around the onset of the linear regime. While the exact MINEP solution displays an exponential dependence on the position, it can be shown that under reasonable experimental conditions the deviations between the rigorous MINEP temperature profiles and the steady profiles are small. In this work we are not interested in the careful identification of the range of validity of Eq.(\ref{eq:linregcon}), but rather in the forthcoming geometrical interpretation of the constraint as a circuitation: So we will always assume that our systems admit a well-defined linear regime.

At a steady state as much heat is absorbed by the colder reservoir, as much has to be poured in by the hotter one. If we ideally short-circuitate the rod, bringing the end-points to coincide, the linear system is mapped to a unicyclic system, with a conserved heat flux through the whole ring. Due to the discontinuity of $T(x)$ at $x=a$, the affinity is not a conservative field (i.e., it is not the derivative of some potential in all of its domain). However, we can still integrate it to get the constraint
\be \oint_{\mathrm{ring}} a(x)dx \; \equiv \; \frac{1}{T_a} - \frac{1}{T_b}, \ee
which in the linear regime is equivalent to Eq.(\ref{eq:cond}). When the boundary temperatures coincide, that is, at equilibrium, the affinity is indeed an exact form and the circulation vanishes. Thus there exists a correspondence between ``topological'' circuitations, nonconservative driving forces, and the onset of nonequilibrium behavior.

Schnakenberg's intuition was that circuitations of nonconservative fields are the fundamental observables that keep a system out of equilibrium. We push this further claiming that, in the linear regime, they are the constraints to be imposed to the MINEP procedure.

As soon as one abandons the 1-dimensional case, one incurs  great difficulties. In particular, steadiness $\partial j^\ast = 0$ does not imply  a uniform current distribution, and one realizes that the problem is of geometrical nature, involving differential forms, topology, etc. However, on a discrete state space Schnakenberg's intuition can be efficiently put to work.

\subsection{\label{schnak}Schnakenberg's theory}

J. Schnakenberg's seminal paper \cite{schnak} is mainly known for the identification of the total entropy production (EP) of a Markovian system, although that element was instrumental --- and actually inessential --- to the formulation of a theory of macroscopic observables as circuitations of local forces.

The theory is synthesized below, starting with a simple example.  We then introduce all the definitions and hypothesis that are strictly necessary to the theory.

\subsubsection{\label{example2} An example}

\begin{figure}
\includegraphics[scale=0.17]{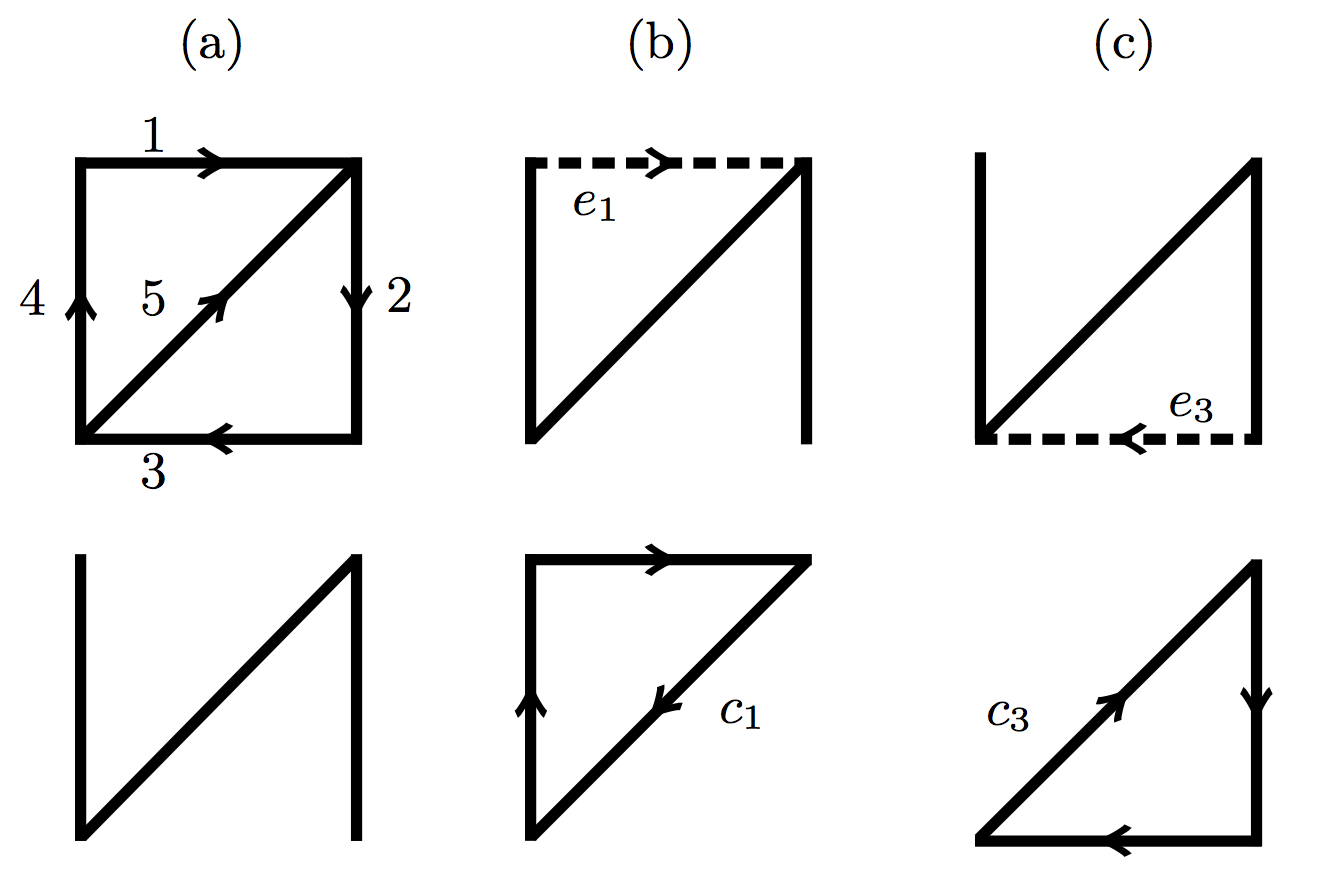}
\caption{\label{fig:cycles}(a) Above: an oriented graph.  Below: a spanning tree. (b) Dotted: a generating chord. Below: its conjugate fundamental cycle. (c) The same.} \end{figure}

Consider a discrete state space consisting of four states, which exchange between one another some physical quantity, be it mass, energy, charge, spin etc., at certain rates. For sake of abstractness, we will suppose that these physical quantities are coded in bits, so that from the comparison of two nearby snapshots of the system an observer will be able to measure a certain flux of raw ``information'' at a certain time, as is shown in Fig.\ref{fig:cycles}a. Here the states of the system are depicted with vertices of a graph, and the channels of communication with oriented edges $e$ connecting the states. Currents $j_e$ might have positive or negative sign, according to the direction of the flow --- concordant or opposite to the edges' orientations. Notice that not all states need to be connected. We further suppose that the currents are induced by some conjugate local forces $a_e$, which have the same sign, and finally we introduce the EP,
\be
\sigma[j,a] = j_1 a_1 + j_2 a_2 + j_3 a_3 + j_4 a_4 + j_5 a_5. \label{eq:epex1}
\ee
A comment is needed on the usage of the scale words. Schnakenberg referred to $j_{e}$ as a microscopic current, and to the observables we are going to build as macroscopic. However, later developments in the stochastic thermodynamics of master equation systems (see \cite{seifert} and references therein) allow us to identify single-trajectory analogs of thermodynamical quantities, such as currents and entropy production, whose averages over paths return $j_e, \sigma,$ etc. This suggests to reserve the word ``microscopic'' for this further layer, and to adopt ``mesoscopic'' for $j_e$ and $a_e$, irregardless of their spatial dimension.

The configuration of currents is \textit{steady} if the total inflow at the nodes is null, yielding the conservation laws
\be
j^\ast_4 = j^\ast_1,\quad j^\ast_2=j^\ast_3,\quad j^\ast_1+j^\ast_5 = j^\ast_2, \quad j^\ast_3-j^\ast_4 = j^\ast_5 . \label{eq:consex} 
\ee
One of them is redundant. The others allow us to express all of the steady currents in terms of, e.g., $j^\ast_1$ and $j^\ast_3$. Replacing the solution in the expression for the EP yields
\be
\sigma[j^\ast,a] =  j^\ast_1 \stackrel{A^1}{\overbrace{(a_1+a_4-a_5)}} + j^\ast_3 \stackrel{A^3}{\overbrace{(a_2+a_3+a_5)}} \label{eq:epex2}.
\ee
Overbraces are used to define the \textit{macroscopic forces} or \textit{affinities}, which are conjugate to the \textit{fundamental currents} $J_1 = j_1^\ast$ and $J_3=j_3^\ast$. The affinities are sums of the local forces along oriented cycles of the graph.

The linear regime constitutive relations are now assumed: currents and forces are related by  $a_e = \ell_e j^\ast_e$, with $\ell_e$ positive local linear response coefficients. We obtain for the macroscopic forces
\begin{subequations}
\bea
A^1 \; = \; \ell_1 j^\ast_1 + \ell_4 j^\ast_4 - \ell_5 j^\ast_5  \; = ~ \stackrel{L^{11}}{\overbrace{(\ell_1+\ell_4+\ell_5)}} J_1 \; + \;  (\stackrel{L^{13}}{\overbrace{-\ell_5}}) J_3 \nonumber  \\ \\
A^3 \; = \; \ell_2 j^\ast_2 + \ell_3 j^\ast_3 + \ell_5 j^\ast_5  \; = ~  \stackrel{L^{33}}{\overbrace{(\ell_2+\ell_3+\ell_5)}} J_3 \; + \; (\stackrel{L^{31}}{\overbrace{-\ell_5}}) J_1 . \nonumber \\
\eea
\end{subequations}

The right-and side defines the macroscopic linear response coefficients, which satisfy Onsager's reciprocity relations.

We gather that the conservation laws at the nodes can be used to express the EP in terms of a certain number of boundary currents and of conjugate affinities, which are circuitations of the local forces along oriented cycles of the graph. Assuming the linear regime constitutive relations yields a symmetric linear response matrix between affinities and fundamental currents.


\subsubsection{Graphs, cycles, fundamental sets}

Consider now a finite microstate space $V$, with $|V|$ microstates labeled by $i,j,\ldots$, which are 
pairwise connected by $|E|$ edges $e$ in the edge set $E$ of a graph $G=(V,E)$. Assign an arbitrary orientation to the edges, which is the choice of tip and tail vertices; $-e$ depicts the inverse edge. The incidence matrix $\partial$, with entries
\be
\partial_i^e =  \left\{ \begin{array}{ll} + 1, &~~ \mathrm{if}~ \stackrel{e}{\gets} i \\ 
-1,  &~~ \mathrm{if}~ \stackrel{e}{\to} i\\
0, & ~~\mathrm{elsewhere}
\end{array} \right. ,
\ee
contains all the information about the topology of the graph, with the exception of irrelevant loops (edges whose tip and tail coincide). Its rows are not independent, as each column adds up to zero. It is a basic graph-theoretical fact that if $\partial$ cannot be arranged in blocks, than it has rank $|V|-1$. The assumption, to which we stick, corresponds to the choice of a connected graph, i.e. a graph whose vertices can all be connected by paths.

From a graphical viewpoint, \textit{cycles} $c$ are chains of edges with no boundary: each vertex it touches is the tip and the tail of an equal number of edges of the cycle. To give an algebraic characterization, notice that, by definition, cycles belong to the kernel of $\partial$; they form the \textit{cycle space} $C$. The dimension of $C$ is called the \textit{cyclomatic number}. By the rank-nullity theorem, the number of independent rows and the dimension of the kernel of a matrix add to the number of its columns, $\mathrm{rk~} \partial + \mathrm{null~} \partial = |E|$. Hence the cyclomatic number is
\be |C| = |E|-|V|+1. \label{eq:cyclomatic} \ee
Cycles are \textit{simple} when they do not have multiple components, overlapping edges nor crossings \footnote{We adopt common jargon in graph theory, whilst Schnakenberg coined the word \textit{quasicycles} for cycles and referred to simple cycles as, simply, cycles.}.

Among all possible basis of $C$, we focus on \textit{fundamental sets}, which are so constructed (see Fig.\ref{fig:cycles}). Let $T \subseteq E$ be a \textit{spanning tree} of the graph, i.e., a maximal set of edges that contains no cycles. Spanning trees enjoy several properties; among others, they touch all the vertices, and consist of $|V|-1$ edges. An edge $e_\alpha$ that does not belong to the spanning tree is called a \textit{chord}. There are $|C|$ chords.
Adding a chord to a spanning tree generates a simple cycle $c^\alpha$, which can be oriented accordingly with the orientation of $e_\alpha$. The  fundamental set of cycles so generated can be proven to be a basis for $C$. The proof roughly goes as follows: Any chord belongs to a cycle. By construction, no two fundamental cycles share a chord, so that they are necessarily independent. Finally, a set of $|C|$ independent vectors in $C$ constitutes a basis.

This is the central technicality behind Schnakenberg's theory. Let us resume: Cycles of a graph form  an integer vector space; they span the kernel of the incidence matrix. Adding chords to a spanning tree  generates a basis of simple cycles $c^\alpha$. The vector representative of a simple cycle has components 
\be
c_e^\alpha = \left\{\ba{ll} +1 , & \mathrm{if}~ e ~\mathrm{belongs~ to ~cycle~} c^\alpha  \\
-1 , & \mathrm{if}~ -e ~\mathrm{belongs~ to ~cycle~} c^\alpha  \\
0 & \mathrm{elsewhere.}
\ea  \right.
\ee
A basis which is built out of a spanning tree is said to be a \textit{fundamental set}. Fundamental sets satisfy two very important properties: i) Each simple cycle comes along with a preferred generating edge $e_\alpha$, called a chord, which belongs to the complement of the tree; ii) Each chord belongs only  to the cycle it generates.

\subsubsection{Macroscopic observables}

Let the \textit{mesoscopic currents} $j_e$ be real edge variables, antisymmetric by edge inversion, $j_{-e} = -j_{e}$. Similarly defined are the \textit{mescocopic forces} $a_e$, which are required to bear the same sign as their conjugate currents $j_e$. The entropy production is the bilinear form
\be
\sigma = \sum_e j_e a_e . \label{eq:EP}
\ee
In general, one will assume that currents and forces are interdependent, in such a way that forces vanish when currents vanish. Local linear regime holds when currents and forces are linearly related in a local way (that is, edge by edge) by
\be
a_e = \ell_e j_e, \label{eq:linreg}
\ee 
where $\ell_e = \ell_{-e}$ is a positive symmetric edge variable. The linear regime holds to first order for small currents and forces. Notice that by reasons of symmetry the next leading order in this expression is third order in the currents.

The collection of currents $j^\ast$ is said to be steady when Kirchhoff's law is satisfied at each node,
\be
\sum_e \partial_i^e j^\ast_e = 0 \label{eq:currentlaw},
\ee
or in matrix notation simply $\partial j^\ast = 0$. Hence $j^\ast$ belongs to the kernel of $\partial$, and can be decomposed in a basis of fundamental cycles
\be
j^\ast = \sum_\alpha J_\alpha c^\alpha \label{eq:current}.
\ee
In particular, since chord $e_\alpha$ belongs only  to cycle $c^\alpha$, the \textit{macroscopic internal current} $J_\alpha$ coincides with the current $j_{e_\alpha}$ flowing  along chord $e_\alpha$. Replacing the solution in the expression for the entropy production yields
\be
\sigma^\ast = \sum_\alpha J^\alpha A^\alpha \label{eq:ssep}
\ee
where
\be A^\alpha = \sum_{e} c_e^\alpha a_e  \label{eq:looplaw} \ee
is a \textit{macroscopic external force} or \textit{affinity}, obtained as the circuitation of the mesoscopic forces along cycle $c^\alpha$. Macroscopic currents and affinities are conjugate variables which completely characterize the steady state. When all affinities vanish, the system is said to satisfy \textit{detailed balance}, or to be equilibrium. In the linear regime, replacing the solution to Kirkhhoff's law (\ref{eq:current}) into Eq.(\ref{eq:linreg}) and summing along oriented edges of a cycle yields
\be
A^\alpha = \sum_\beta J_\beta \sum_{e} \ell_e c_e^\alpha c_e^\beta =  \sum_\beta L^{\alpha\beta} J_\beta \label{eq:linresmac} 
\ee
where the \textit{linear response matrix} $L^{\alpha\beta} =\sum_e \ell_e c_e^\alpha c_e^\beta$, being manifestly symmetric, satisfies Onsager's reciprocity relations. This was one main clue that led Schnakenberg to promote macroscopic currents and forces to fundamental observables.

The linear response matrix can be combinatorially constructed by taking all response coefficients $\ell_e$ that belong to both cycles $c^\alpha$ and $c^\beta$, with a plus or minus sign whether the cycles' orientation is either concordant or opposite along edge $e$. This kind of matrices is well known in graph theory and in Feynman diagrammatics \cite{nakanishi}. One property that is relevant to our discourse is that its determinant is always non-null (but for very trivial graphs), which guarantees invertibility of expression (\ref{eq:linresmac}).

\subsection{\label{MINEP} Minimum entropy production principle}

In this section we prove that, in the linear regime, steady states minimize entropy production among all states that are compatible with the external macroscopic circuitations of the mesoscopic forces. We then specialize the result to master equation systems, showing that in the linear regime the steady probability distribution minimizes the entropy production.

\subsubsection{\label{general} General statement}

According to Schnakenberg's theory of nonequilibrium observables, the external constraints that force a system into a non-equilibrium steady state are the macroscopic external forces. We will now extremize entropy production with respect to mesoscopic currents in the linear regime, using Schnakenberg's affinities as constraints,
\be \bar{A}^\alpha \equiv A^\alpha[j] = \sum_{e}  c^\alpha_e \ell_e j_e + O(j^3),\label{eq:constraints} \ee
where $\bar{A}^\alpha$ is a fixed, ``observed'' value of the affinity. The EP is the quadratic form
\be \sigma[j] = \sum_{e} \ell_e j_e^2 + O(j^4). \ee
We introduce Lagrange multipliers and vary
\be \frac{\delta}{\delta j_e}\left[\sigma[j] - 2\sum_\alpha \lambda_\alpha \Big( A^\alpha[j] - \bar{A}^\alpha \Big) \right] = 0. \label{eq:variation} \ee
Multipliers $\lambda_\alpha$ are to be determined by replacement of the stationary solution into (\ref{eq:constraints}). The calculation is easily carried over, yielding
\be j^\ast = \sum_{\alpha} \lambda_\alpha c^\alpha . \ee
Stationary currents are linear combinations of a cyclomatic number of boundary terms $\lambda_\alpha$. We now prove that the latter are in fact the macroscopic currents conjugate to the constrained values of the affinities. Fixing the Lagrange multipliers we get
\be \bar{A}^\alpha = \sum_{\beta} \lambda_{\beta} \sum_{e} \ell_e c_e^\alpha c_e^\beta = \sum_{\beta} L^{\alpha\beta} \lambda_{\beta},  \ee
where we recognized the linear response matrix. This relation can be inverted, showing, after Eq.({\ref{eq:linresmac}), that $\lambda_\alpha$ is the steady current conjugate to the measured value of the affinity $\bar{A}^\alpha$.
The second variation
\be \frac{\delta \sigma}{\delta j_e \delta j_f} - 2 \sum_{\alpha} c_e^\alpha \frac{\delta  \lambda_\alpha }{\delta j_f}=  \ell_e \delta_{e,f}.  \ee
yields a positive Hessian matrix, which guarantees positive concavity.

We thus conclude that the stationary configuration of currents that in the linear regime minimizes the entropy production with constrained macroscopic forces, satisfies Kirchhoff's conservation law. From a dynamical point of view, if under some ergodic hypothesis the steady state is asymptotically reached over the long time (as is the case, for example, for Markovian systems), one can conclude that a nonequilibrium system tends to relax to a state of minimum entropy production, compatibly with the macroscopic external forces that prevent it from reaching equilibrium. This echoes Klein and Meijer's phrasing.
 
\subsubsection{\label{master}Master equation systems}

Up to this point our main variables have been the currents, whose nature and origin was left unspecified. When dealing with Markov processes the object of interest is a normalized probability distribution $\rho$ on the vertex set, which we will call a \textit{state} of the system. Probability currents are defined in terms of it as
\be j_{ij}[\rho] = w_{ij} \rho_j - w_{ji} \rho_i, \ee
where the $w_{ij}$'s are positive time-independent transition rates between states, with the physical dimension of an inverse time. The master equation dictates the evolution
\be
\frac{d}{dt} \rho(t) = L \rho(t) = - \partial j[\rho(t)],
\ee
where $L$ is  the Markovian generator. Its solution is $\rho(t) = e^{t L} \rho_0$ for any initial state $\rho_0$. 
We assume that the graph $G$ is connected (that is, that there exists a path between any two vertices) and that transition rates are non-null in both directions along its edges. Under these mild assumptions, there exists a unique invariant, or steady, state $\mu$, $L \mu = 0$, which is approached at late times,
\be \rho(t) ~\stackrel{t\to \infty}\longrightarrow ~ \mu, \quad \forall \rho_0,   \ee
independently of the initial distribution. Notice that the invariant state is uniquely determined from knowledge of the system's transition rates, and in fact there exists an explicit (but complicated) combinatorial expression for it, which is unnecessary for the present discussion \cite{schnak}.

The conjugate force along edge $e= i\gets j$  is defined as
\be a_{ij}[\rho] = \ln \frac{w_{ij} \rho_j}{w_{ji}\rho_i}, \label{eq:defaff}  \ee
and it is dimensionless. By the Handshaking Lemma we can replace the sum over edges with one-half the sum over neighboring sites, yielding the EP
\be
\sigma[\rho] = \tfrac{1}{2} \sum_{i,j} j_{ij}[\rho] a_{ij}[\rho] = \tfrac{d}{dt} S[\rho] + \sigma_\mathrm{env}[\rho] . \label{eq:schEP}  \ee
The motivation behind definition (\ref{eq:defaff}), also due to Schnakenberg \cite{schnak}, is that the entropy production  is naturally split in the time derivative of the Gibbs-Shannon internal entropy $S[\rho]$, and a term that quantifies heat exchange with the environment, yielding an entropic balance equation that is generally widely accepted, as it is the mesoscopic counterpart of the first law of (stochastic) thermodynamics along single trajectories \cite{seifert}.

The main peculiarity of Def.(\ref{eq:defaff}) is that  affinities turn out to be independent of the state of the system $\rho(t)$, and thus time-independent,
\be
A^\alpha = \ln \prod_{e \in c^\alpha} \frac{w_{e}}{w_{-e}}.
\ee
This property was another important clue that led Schnakenberg to interpret them as environmental constraints. Remarkably, this further entails that, for a fixed set of transition rates, variation of the probability distribution $\rho$ in any case preserves all of the macroscopic affinities: We do not need to impose any constraint at all, as long as we stick to the linear regime.

Let us then set up the linear regime. Consider a set of equilibrium transition rates $w_{ij}^0$, whose steady state $\mu^0$ satisfies detailed balance
\be \frac{w^0_{ij} \mu^0_j}{w^0_{ji} \mu^0_i} = 1, \label{eq:detbal} \ee
which by Kolmogorov's criterion is known to hold if and only if all of the affinities vanish. The linear regime is attained when perturbing the equilibrium generator to a non-equilibrium one, $L^\epsilon$, and at the same time considering a probability distribution $\rho^\epsilon$ which is only slightly apart from the equilibrium steady state,
\begin{subequations}
\bea
w^\epsilon_{ij} & = &   \left( 1 + \varepsilon_{ij} \right) w^0_{ij}  \\
\rho^\epsilon_i & = & \left(1 + \eta_i\right)  \mu^0_i, \eea
\end{subequations} 
where we suppose that  all the $\varepsilon$'s and $ \eta$'s are of infinitesimal order $\epsilon$. Let $\mu^\epsilon$ be the steady state relative to generator $L^\epsilon$. It can be shown \cite{maes1} that $\mu^\epsilon - \mu^0$ is of order $\epsilon$, and consequently so is $\rho^\epsilon - \mu^\epsilon$.
Expanding the forces and the currents we obtain
\begin{subequations}
\bea
 a_{ij}[\rho^\epsilon]Ê& = & \ln \left( 1 + \frac{w^\epsilon_{ij}\rho^\epsilon_j - w^\epsilon_{ji} \rho^\epsilon_i}{w^\epsilon_{ji}\rho^\epsilon_i} \right)Ê ~\approx~ \frac{j_{ij}[\rho^\epsilon]}{w^0_{ji} \rho^0_i} \label{eq:response} \\
j_{ij}[\rho^\epsilon] & =  & w^0_{ij} \rho^0_j  \Big( \varepsilon_{ij} + \eta_j - \varepsilon_{ji} - \eta_i \Big). 
 \eea
 \end{subequations}
The r.h.s. of Eq.(\ref{eq:response}) furnishes the linear response coefficient, $\ell_{ij} = (w^0_{ji} \rho^0_i)^{-1}$, which by Eq.(\ref{eq:detbal}) is indeed symmetrical under edge inversion. Notice that both forces and currents are of order $\epsilon$; entropy production is of order $\epsilon^2$.

In the linear regime local forces and currents meet all the requirements that are necessary to formulate the MINEP principle proven above. On the other hand, macroscopic constraints are independent of the probability distribution. To conclude, we need to prove that the following variational problems are equivalent:
\be
\left. \frac{\delta \sigma}{\delta j} \right|_{A^\alpha} = 0 \; \Leftrightarrow \; \frac{\delta \sigma}{\delta \rho} = 0.
\label{eq:implication} \ee
Using Eq.(\ref{eq:schEP}) and Eq.(\ref{eq:response}), we can write the variation with respect to $\rho_k$ as a linear combination of the variations with respect to the currents:
\be
\frac{ \delta \sigma[\rho]}{\delta \rho_k} =   \sum_{i} w_{ik}\frac{\delta \sigma[j]}{\delta j_{ik}}.
\ee
Since  variation of $\rho$ preserves the affinities, we also have
\be
  \left. \sum_{i} w_{ik}\frac{\delta \sigma[j]}{\delta j_{ik}}\right|_{A^\alpha} =  \left. \frac{ \delta \sigma[\rho]}{\delta \rho_k} \right|_{A^\alpha}=  \frac{ \delta \sigma[\rho]}{\delta \rho_k}.
\ee
Whent the r.h.s. vanishes, so does the l.h.s.. Hence the left-to-right implication in Eq.(\ref{eq:implication}) can be drawn, which suffices to prove that the MINEP principle for master equation systems follows from ours: the invariant state $\mu^\epsilon$ is a local minimum of the entropy production among nearby probability distributions. Notice that we cannot continue our conclusion where $\eta$ is no longer small. So we might expect that, even for near-equilibrium transition rates, there might exist a landscape of minima of the entropy production besides the invariant state.

For sake of completeness, notice that the inverse implication in (\ref{eq:implication}) also holds. We need to make sure that the linear span of the variables is the same, in order to avoid, for example, that negative curvature directions of possible saddle points be out of the span of the probability distribution, but within the span of the currents. In such a case, one would conclude that certain configurations of currents are extremal without being able to inspect all  possible configurations. A qualitative argument goes as follows. There are $|E|$ currents, subject to a cyclomatic number of linear constraints, and $|V|$ probability entries, subject to one linear constraint, namely normalization. From  Eq.(\ref{eq:cyclomatic}) and from the fact that currents are linear in the probabilities, it follows that the linear span of the two variations has the same dimensionality: Hence they cover the same small neighborhood in the space of current profiles, near the equilibrium steady state. There is one last subtlety involved with this: Probability densities live in a simplex $0 \leq \rho_i \leq 1$, rather than a vector space, which for small-enough rates would make even small currents out of reach. This is not a true limitation, as transition rates are defined up to the conventional choice of an unit measure of time, which of course does not affect the generality of the principle \footnote{In other words, the question arises: with respect to \textit{what} are current small? A better definition of linear regime is then the requirement that dimensionless forces, and not currents, be small; in that case inspection of Eq. (\ref{eq:response}) reveals that $\rho_i \in [0,1]$ poses no limitations.}. 

\subsection{\label{relationship}Discussion: relationship to previous results}

The first  paragraph  of this section is devoted to tributing Prigogine's insights. Our proof of MINEP for master equation systems is a new path to an old result; we then briefly credit the two approaches we are aware of.  We conclude with a discussion of a formulation of the maximum entropy production principle that also involves Kirchhoff's equations.

\subsubsection{\label{prigo} Prigogine}

I. Prigogine's proof of MINEP \cite{prigo}, shaped upon chemical systems, is based on an assumed splitting of the EP into a matter flux term and a heat flux term,
\be
\sigma = J_{th} A_{th} + J_m A_m .
\ee
Steadiness is equivalent to the requirement that matter currents vanish, $J_m = 0$. Hence EP at a steady state consists only of heat flux contributions. 

Bridging to our abstract setup, we might interpret heat currents and inverse temperature gradients respectively as Schnakenberg's internal currents and external forces (this is precisely the case in the example discussed in SubSec. \ref{example}). Pushing this identification out of the steady state, the question arises whether there exist conjugate 
observables analogous to ``matter currents'' and ``pressure gradients'' that would allow an analogous splitting in terms of Schnakenberg-type observables. The answer is in the affirmative. In \cite{polettini}, the author provides a complete set of observables, complementing Schnakenberg's with \textit{internal forces} $A_\mu$  and \textit{external currents} $J^\mu$, in such a way to bring EP to the form
\be
\sigma = \sum_\alpha A^\alpha J_\alpha + \sum_\mu J^\mu A_\mu . \label{eq:rearr} 
\ee
The new observables are again built as linear combinations of mesoscopic observables along certain subgraphs. External currents vanish at a steady state.

A review of the construction and its detailed properties goes beyond the scope of this article. Let us hint at it by finalizing the special case treated in SubSec. \ref{example2}. Basically, we want to rearrange Eq.(\ref{eq:epex1}) so to have cycles emerge. Since the new expression should reduce to Eq.(\ref{eq:epex2}) when the conservation laws (\ref{eq:consex}) hold, regardless of the value of the local affinities, we are indeed able to collect the currents in a profitable way:
\bea
\sigma[j,a] ~=~  j_1 \stackrel{A^1}{\overbrace{(a_1+a_4-a_5)}} ~+~ j_3 \stackrel{A^3}{\overbrace{(a_2+a_3+a_5)}} \nonumber \\
 +~ a_2 \stackrel{J^2}{\overbrace{(j_2-j_3)}} ~+~ a_4 \stackrel{J^4}{\overbrace{(j_4-j_1)}} ~+~ a_5 \stackrel{J^5}{\overbrace{(j_5 + j_1 - j_3)}}
 .
\eea
The \textit{macroscopic external currents} $J^2,J^4,J^5$ vanish at the steady state, as does $J_m$ in Prigogine's approach. Notice that index  $\alpha$ in Eq.(\ref{eq:rearr}) ranges over $1,3$, index $\mu$ ranges over $2,4,5$, and that the alternating index positioning is crucial to the identification of the diverse observables. Exploiting the linear regime local constitutive relations, one obtains
\begin{subequations}\label{eq:linmac}
\bea
J_\alpha = \sum_\beta \Gamma_{\alpha\beta} A^\beta + \sum_\mu \Gamma_\alpha^\mu A_\mu \\
J^\mu = \sum_\alpha \Gamma_{\alpha}^\mu A^\alpha + \sum_\nu\Gamma^{\mu\nu} A_\nu ,
\eea
\end{subequations} 
which pairs with Prigogine's Eq.(6.2). Direct calculation of the coefficients (which we leave to the reader) shows that the reciprocity relations are fulfilled.  Now the reasoning follows along the same tracks as Prigogine's. Variation with respect to the mesoscopic observables, at constant external affinities, can be replaced with variation with respect to the internal forces
\be
\left. \frac{\delta}{\delta a}\right|_{A^\alpha} \; \Leftrightarrow \; \frac{\delta}{\delta A_\mu}.
\ee
Replacing Eq.s (\ref{eq:linmac}) into Eq. (\ref{eq:rearr}) and varying,  
\be
 \frac{\delta \sigma}{\delta A_\mu} = 2 (\Gamma^\mu_\alpha A^\alpha + \Gamma^{\mu\nu} A_\nu ) = 2J^\mu = 0,
\ee
yields vanishing external currents. We thus conclude that our approach is completely superimposable on Prigogine's phenomenological derivation, adding to it an abstract and quite general definition of the constraints.

\subsubsection{Jiu-Li  and coworkers, Maes and coworkers}

The problem of extending Prigogine's theorem to a statement regarding populations, probability distributions or density matrices was raised and undertaken already by Klein and Meijer \cite{klein} with a specific model, and by Callen \cite{callen} in a quantum mechanical setting, in which a number of fixed forces are assumed [see Eq. (39)]. 

Later, the Brussel school delved into the problem of establishing stability criteria for nonequilibrium steady states \cite[Sec.3.5]{glan}. Along this line of research, Schnakenberg's expression (\ref{eq:schEP}) for the EP of a Markovian system first appears in the work of L. Jiu-Li, C. Van den Broeck and G. Nicolis \cite{luo}, who derived the MINEP principle for probability distributions evolving under a master equation in a very straightforward manner, which we now synthesize. The time derivative of a local force is
\be \dot{a}_{ij}[\rho] = \frac{\dot{\rho}_j}{\rho_j} - \frac{\dot{\rho}_i}{\rho_i}.  \ee
Writing the linear regime expression for the entropy production in terms of the affinities
\be
\sigma = \tfrac{1}{2} \sum_{i,j} a_{ij}^2 / \ell_{ij}
\ee
and taking its time derivative, we obtain
\be
\dot{\sigma} = 2 \sum_{i,j} \frac{a_{ij}}{\ell_{ij}}  \frac{\dot{\rho}_j}{\rho_j} = - 2\sum_{i,j} j_{ij} \frac{\dot{\rho}_i}{\rho_i}  = - 2\sum_{i}  \frac{(\dot{\rho}_i)^2}{\rho_i}.\ee
Since $\dot{\sigma} \leq 0$, the EP decreases in the vicinity of the steady state toward the steady state.

More involved is the approach of Maes and Neto\v cn\'y \cite{maes1}, who considered the large deviation rate function $I^\epsilon$ of the occupancy empirical distribution
\be I^\epsilon[\rho] = - \lim_{T\to \infty}  \frac{1}{T} \log P\left( \frac{1}{T} \int_0^T \delta_{i,\xi_t} dt \equiv \rho_i\right).\ee
Here $\xi_t$ is a single jump-process trajectory from time $0$ to time $T$, and $P$ is the probability measure over trajectories.  Maes and Neto\v cn\'y  proved the stronger result that, near equilibrium, in the leading order the rate function $I^\epsilon[\rho]$ is equal to one-fourth the entropy production difference $\sigma[\rho] - \sigma[\mu^\epsilon]$ between state $\rho$ and the invariant state. Since upon the above assumptions, on connectedness and non-vanishing rates, Markovian systems are known to converge
to the steady state, and since, by the law of large numbers, the steady state $\mu^\epsilon$ is a global minimum of the rate function, one concludes that it is a local minimum of the entropy production.

\subsubsection{ \label{MAXEP} \v Zupanovi\'c and coworkers}

The striving for variational principles in NESM has a long and contrived history. In particular, another, less familiar, variational principle has been proposed that should characterize the behavior of non-equilibrium systems: the \textit{maximum entropy production principle} (MAXEP). There are at least as many formulations of MAXEP as there are of MINEP. Arguably, the apparent clash between these two instances is due to the fact that they apply to distinct scales and regimes, and employ different notions of ``state''. There is a vast literature that tries to sort out the matter
\cite{martyu}, and by no means do we mean to be exhaustive. However, we need to put our principle in contact with some instances of MAXEP in order to appreciate their relative significance.

It was Jaynes's conviction that \cite{jaynes} \textquotedblleft there must exist an exact variational principle for steady irreversible processes\textquotedblright
and that such  principle should capture conservation laws:
\textquotedblleft we should rather take the conservation laws as exact and given, and seek a principle which gives the correct phenomenological relations\textquotedblright.
Jaynes thought that reversing this logic would also reverse the principle:
\textquotedblleft perhaps the exact phenomenology is the one that has maximum entropy production for prescribed exact conservation laws\textquotedblright.
So, Jaynes's expectation was that conservation laws and constitutive equations should fit in the same picture, under the aegis of one unifying maximum principle. This supposition informs Gyarmati's research \cite[Par. V.3]{gyarmati}, as he claims that \textquotedblleft the principle of minimum production of entropy is not an independent principle [\ldots], but rather is only an alternative reformulation of the Onsager principle valid for stationary cases\textquotedblright.

In this respect, P. \v Zupanovi\'c, D. Jureti\'c and S. Botri\'c's proposition is more closely related to our principle, as it deals with Kirkhhoff's current law  (\ref{eq:currentlaw}) and Kirkchhoff's loop law (\ref{eq:looplaw}) on networks. It is to this work that we mainly refer in the following.

Suppose we do not know the system's constitutive equations, but that we do know that entropy (in the form of heat) is dispersed into the environment at a rate
\be \omega = \sum_e \ell_e j_e^2, \ee
which is called the \textit{dissipation function}. This is the case for electrical circuits, where $\ell_e$ plays the role of a resistance. Entropic balance then requires
\be \sigma - \omega \equiv 0. \label{eq:constrmax} \ee
This is particularly reasonable for an electrical circuit, where $\omega$ is the electric power and $\sigma$ the heat flux \cite{looplaw}. Finally we extremize entropy production, varying with respect to the currents, and imposing constraint (\ref{eq:constrmax}),
\be \frac{\delta}{\delta j_e} \Big[\sigma + \lambda (\sigma - \omega) \Big] = 0. \label{eq:variation} \ee
We obtain
$a_e= 2 \lambda / (1+\lambda)  ~ \ell_e j_e. $
The value of the multiplier is set by replacing the extremizer in Eq.(\ref{eq:constrmax}), which yields $\lambda = 1$, a stationary value $\sigma^\ast = \omega$ and the desired mesoscopic phenomenological, $a_e = \ell_e j_e$. Taking the second variation we obtain a negative hessian, hence a concave-down paraboloid, hence we front a maximum entropy production principle.

Variational principle (\ref{eq:variation})  is discussed by Martyusheva and Seleznev \cite[Eq. 1.16]{martyu}, where it is introduced as Ziegler's principle, and again by \v Zupanovi\'c and coworkers \cite[Eq. (9)]{zupan} in a follow-up paper on the relation between MAXEP and the principle of least dissipation: In fact,  the procedure is but a restatement of Onsager's least dissipation principle, which in its original form simply states that $\sigma-\omega$ should be maximum \cite{onsager}.

Embedding Kirchhoff's current law (\ref{eq:currentlaw}) into Eq.(\ref{eq:variation}),
\be \frac{\delta}{\delta J_\gamma} \Big[(1+\lambda) \sum_\alpha J_\alpha A^\alpha - \lambda \sum_{\alpha,\beta} L^{\alpha\beta} J_\alpha J_\beta  \Big] = 0,  \ee
or, equivalently, constraining the solution to the variational problem (\ref{eq:variation}) on the $\partial j = 0$ shell, leads to the identification of circuitations $A^\alpha = \sum_{\beta} L^{\alpha\beta} J_\beta$ as the phenomenological conjugate variables to the currents. This realization of MAXEP does indeed reproduce Jaynes's expectation that the reversed logic should yield the correct phenomenological laws. The MAXEP of \v Zupanovi\'c \textit{et al.} is in a sense complementary to our MINEP, reproducing the macroscopic Onsager's relations. With one specification: The MAXEP principle does \textit{not} imply that  \textquotedblleft currents in a linear planar network arrange themselves so as to achieve the state of maximum entropy production\textquotedblright. That is due to the minimum entropy production principle.

\subsection{ \label{concl}Concluding remarks}

Some of the hypothesis upon which we derived the principle can be relaxed. Working in a differential-geometric setting should allow us to extend the principle to continuous systems. In this context, a result similar to Schnakenberg's decomposition has been obtained by Jiang and the Qians \cite{qiansbook} for topological currents,  such as those that flow along the two fundamental cycles of a torus. The task is then to extend to nontopological currents, through lattice discretizations and limiting procedures. One problem appears at the horizon: As the discretization becomes more and more refined, the number of cycles tends to infinite, becoming non-denumerable in the continuum limit. This clashes with the physical intuition that nonequilibrium constraints should be a few boundary conditions that are experimentally accessible. For physically relevant systems, symmetries might have a role in the reduction of the number of affinities.

The condition of locality can also be relaxed, considering the more general mesoscopic phenomenological linear response relation
\be a_e = \sum_f \ell_{ef} j_f, \label{eq:nonloc} \ee
where $(\ell_{ef})_{e,f}$ is required to be a positive symmetric matrix. Definitions and proofs become only slighly more complicated by considering (\ref{eq:nonloc}) in place of (\ref{eq:linreg}), with the only exception of $\det L \neq 0$, whose proof might be nontrivial.

As to the hypothesis of linear regime, in our formulation the assumption seems to be unavoidable if one chooses affinities as nonequilibrium constraints. The possibility is open that better observables might allow for a departure from the linear regime. There exist many instances of variational principles in NESM, most of which can be traced back to Onsager's least dissipation ad Prigogine's minimum EP, with their own, and different, inclination. 
However, 
to the author's knowledge none truly departs from the linear regime, at least in an operational sense. So, in this respect our principle is no exception.

Our extremization procedure is based on the identification of the fundamental macroscopic observables which keep a system in a nonequilibrium steady state. The setup is quite general and can in principle be adapted to any system that allows a local conservation law. It can be applied to master equation systems, where its robustness can be tested against well-known results. It is shown to provide an abstract setup where Prigogine's original statement sits comfortably, provided that we have a mesoscopic substrate. So, while the principle \textit{per se} is no novelty, the procedure and its generality are.

While the hot topic of NESM are, of course, fluctuations, there is \textit{a priori} no fluctuating character in the principle we have formulated: It is purely geometrical, as one can show by recasting the construction in the language of discrete differential forms \cite{ddf}.

Coming to a conclusion, we suggest that the search for an extremal functional is as important as the identification of constraints of physical relevance. This might be a good guiding principle, for example, in the search of a maximum entropy principle: While MAXENT can be constructively employed to derive equilibrium ensembles \cite{maxent1}, to our knowledge a similar application to nonequilibrium steady states is still lacking. The possibility is open that giving to Schnakenberg's affinities the correct weight might allow to derive as useful tools of calculation as are equilibrium ensembles, fulfilling Jaynes's expectation that \textquotedblleft essentially all of the known results of Statistical Mechanics, equilibrium and nonequilibrium, are derivable consequences of this principle\textquotedblright  \cite{jaynes2}.

\paragraph*{Aknowledgments.} The author is grateful to A. Maritan for assistance, and to M. Esposito and F. Sbravati for fruitful discussions.

\end{document}